\documentclass{article}

\usepackage{amssymb}
\usepackage{amsthm}
\usepackage[cmex10]{amsmath}
\usepackage{subeqnarray}
\usepackage{graphicx}
\usepackage{comment}

\begin{document}
\title{Fourth-order flows in surface modelling}

\author{Ty Kang}

\maketitle
\begin{abstract}
This short article is a brief account of the usage of fourth-order curvature
flow in surface modelling.
\end{abstract}

\section{Introduction}

Recent developments on so-called $G^1$ surface modelling have been investigated
\cite{GQ6} with the aid of fourth order geometric flows.
In particular, the surface diffusion flow and the Willmore flow have found
extensive application.
These equations are geometric, in the sense that they do not depend on the
choice of coordinates.
Classical problems to which these can be applied include surface blending,
$N$-sized hole filling, and free-form surface fitting. This last application is
where the $G^1$ boundary conditions arise.

Earlier work has used the mean curvature flow to attack problems related to
those above. The mean curvature flow is a quasilinear second order geometric
evolution equation, which means again that the equation is invariant under
choice of coordinates. The mean curvature flow is the steepest descent
$L^2$-gradient flow of the area functional, and has been investigated
analytically at least since Mullins \cite{Mull1}. A landmark result is that of
Huisken \cite{H84}. Applications of the mean curvature flow to smooth or fair
noisy images is very efficient -- see \cite{10fromGQ6,35fromGQ6} for example. 

Fourth order curvature flows and geometric differential operators have also been
investigated for some time, at least since Mullins \cite{Mull2}, who proposed
the surface diffusion flow. Analytic results on the surface diffusion flow came
quickly. We just mention here \cite{W11slt,W11sdlte,MWW11clt} and the references
contained therein. The Willmore flow was only proposed relatively recently
\cite{KS02} but has received a lot of attention in computer graphics.  Other
fourth order geometric operators such as the biharmonic operator have also been
investigated \cite{6fromGQ6,7fromGQ6}. We have included further references to
the relevant literature in the bibliography.

\section{The Models}

We consider three models: the surface diffusion flow, the Willmore flow, and the
quasi-surface diffusion or `na\'ive biharmonic' flow \cite{46fromGQ6,47fromGQ6}.
This last one is especially interesting as it has promise in surface modelling
\cite{GQ6} but the theoretical analysis is still wide open.

We do not survey the literature any more here and refer to the introduction for
relevant papers, along with the references contained within those.

\subsection{Surface diffusion flow}

The surface diffusion flow is the natural fourth-order version of mean curvature
flow. It is a family $\{M(t)\}_{t>0}$ of closed surfaces evolving according to
\[
\frac{\partial \rho}{\partial t} = - (\Delta H)n
\]
\[
M(0) = M_0
\]
\[
\partial M(t) = \Gamma
\]
where $\Delta$ is the Laplace-Beltrami operator on $M(t)$ and $H$, $n$ are the
mean curvature and the surface normal of $M(t)$ respectively. We note that
\[
\frac{d}{dt} \text{Area }(M(t)) = - \int_{M(t)} |\nabla H|^2dS
\]
where $\nabla$ is the connection on $M(t)$ and $dS$ is the surface measure. Also
\[
\frac{d}{dt} \text{Vol }(M(t)) = 0.
\]
These properties are interesting for surface modelling.

\subsection{Willmore flow}

The Willmore flow is the gradient flow of the $L^2$-norm of the mean curvature
squared. It is a family of surfaces evolving according to
\[
\frac{\partial \rho}{\partial t} = - (\Delta H + 2H(H^2-K))n
\]
\[
M(0) = M_0
\]
\[
\partial M(t) = \Gamma
\]
where $K = \text{det }A$ and $A$ is the second fundamental form, is the Gauss
curvature.

For surface modelling, it is important that the Willmore flow reduces the
average of the curvature across the whole surface very quickly. This makes it
almost ideal for smoothing purposes, where we consider the flow for a very short
time only.

\subsection{Quasi surface diffusion or biharmonic flow}

The (QSDF) is given by
\[
\frac{\partial \rho}{\partial t} = - (\Delta^2 \rho)
\]
\[
M(0) = M_0
\]
\[
\partial M(t) = \Gamma
\]
where $\Delta$ is the Laplace-Beltrami operator on $M(t)$ and $H$, $n$ are the
mean curvature and the surface normal of $M(t)$ respectively. We note that this
flow is not volume preserving and is also not reducing surface area quickly like
the standard surface diffusion flow.

However the interesting property of (QSDF) is that in \cite{GQ6} it is found to
be very effective in $G^1$ surface modelling. Unfortunately theoretical analysis
is not as abundant for this flow as it is for the Willmore and surface diffusion
flows.

\section{Application of the flows}

Let us briefly discuss the application of the (SDF), (WF), and (QSDF) to the
problem of smoothing noisy surfaces.

We first note that the short time behaviour of the flows is very similar, so for
the smoothing of small regions of rough surfaces they appear to be of equal
value.
It is the long term evolution which differentiates the three flows rather
strikingly.
The (WF) reduces curvature quickly and tends to avoid breaking apart the
evolving surface, while (SDF) appears to be happy to tear off various pieces.

The (QSDF) is similar to (SDF), except that it does not respect the initial
shape so closely. This is because while (SDF) preserves volume, the (QSDF) does
not. On the other hand, the (QSDF) is sometimes better behaved, as it reduces
the harmonic energy field (also known as the tension field) of $\rho$ very
quickly.

\subsection{$G^1$ character of the flows}

Each of the (SDF), (QSDF) and (WF) are fourth-order. This means that for the
$G^1$ surface modelling they are ideal and can ensure continuity of the tangent
vectors at the polygon boundary.

This is obviously much better for applications than the mean curvature flow or
other second order geometric pdes, as they can only ensure $G^0$ surface
modelling.

We plan on investigating even higher order geometric PDEs in the future. These
are particularly relevant for modelling of plates in automobiles. The reflective
character of the body is determined by the $G^2$ nature of the plates. If this
is not respected then reflections can flip at polygonal boundaries. This
requries at least geometric PDE of sixth order.

\section*{Acknowledgements}

This work is in progress and forms part of the author's undergraduate thesis.

\nocite{*}
\bibliographystyle{plain}
\bibliography{wheeler,tk}

\end{document}